\documentclass[
  reprint,
  amsmath,amssymb,
  aps,pra,
  superscriptaddress,
  nofootinbib,
  floatfix
]{revtex4-2}

\usepackage{graphicx}
\usepackage{hyperref}
\usepackage{bm}
\usepackage{physics}
\usepackage{xcolor}
\usepackage{siunitx}
\usepackage{braket}
\usepackage[normalem]{ulem}

\newcommand{\dtv}{\delta_{\mathrm{TV}}}
\newcommand{\lmax}{\lambda_{\max}}
\newcommand{\dint}{d_{\mathrm{int}}}

\begin{document}

\title{Comparing Classical Simulation and Sample-Based Learning of Quantum Systems}

\author{Jo\~ao Pedro d'El-Rey}
\email{jpedrodelrey@gmail.com}
\affiliation{Centro Brasileiro de Pesquisas Físicas, Rua Dr. Xavier Sigaud, 150, Rio de Janeiro, RJ, Brazil}
\author{Ra\'ul O. Vallejos}
\affiliation{Centro Brasileiro de Pesquisas Físicas, Rua Dr. Xavier Sigaud, 150, Rio de Janeiro, RJ, Brazil}
\author{Fernando de Melo}
\affiliation{Centro Brasileiro de Pesquisas Físicas, Rua Dr. Xavier Sigaud, 150, Rio de Janeiro, RJ, Brazil}             
\date{\today}

\begin{abstract} 
We investigate the relationship between two distinct classical approaches to quantum systems: direct simulation from a classical description and sample-based learning from measurement data. While both tasks ultimately aim to reproduce Born-rule statistics, complexity-theoretic results suggest that simulability and learnability need not coincide in general. Here we study this relationship empirically using a fixed deep energy-based generative model trained on measurement samples from controlled families of quantum states. We independently tune two quantum resources associated with classical simulation cost: entanglement, through the bond dimension of random matrix product states, and non-stabilizerness, through the number of T gates in Clifford-dominated circuits. Learning difficulty is characterized using two probes of neural-network complexity: the largest Hessian eigenvalue at convergence and Random Subspace Optimization. For both quantum resources, increasing simulation hardness systematically correlates with sharper loss landscapes and degraded reconstruction performance under constrained capacity. Our results indicate that, within the regimes studied here, classical learnability tracks known simulation complexity measures, suggesting that neural-network training dynamics can provide an empirical probe of quantum computational hardness.
\end{abstract}

\maketitle

% ============================================================
\section{Introduction}\label{sec:intro}
% ============================================================

Predicting the behavior of a quantum system from a classical description and reconstructing such behavior from measurement data are two different computational tasks. The first, \emph{simulation}, receives a Hamiltonian or a circuit and outputs predictions; under standard complexity-theoretic assumptions, it is believed to require exponential classical resources in general \cite{bremner2011classical, aaronson2011computational}. The second, \emph{learning}, receives only measurement samples from an unknown state and must produce a predictor that reproduces the measurement statistics \cite{aaronson2007learnability, hinsche2023}. Both paradigms can be evaluated on the same task, namely sampling from or evaluating Born distributions, which makes a direct complexity comparison meaningful.

A natural conjecture is that the two notions of complexity coincide: states that are hard to simulate should also be hard to learn from measurements, and vice versa. Recent results both support and challenge this view. On the supporting side, work on neural-network quantum state tomography has repeatedly identified entanglement, criticality, and system size as bottlenecks for classical learning \cite{torlai2018neural, carrasquilla2017machine, carrasquilla2020machine, du2025artificial}.  On the challenging side, theoretical constructions have exhibited explicit separations: classically simulatable circuits that are hard to learn from samples \cite{hinsche2023, bittel2025pac}, and systems where a hybrid learning+simulation approach is necessary beyond usage of a unique paradigm \cite{huang2022provably}, and proposals that identifies the necessary structure to demonstrate such connections \cite{yoganathan2019}. These results suggest that simulability and learnability may, in general, capture orthogonal facets of quantum complexity.

Even though the theoretical picture is nuanced, neural networks have empirically been shown to respect known complexity transitions \cite{ippoliti2024learnability, kim2025learning}. In these settings, exponential learnability bounds imparted by properties of matter -- previously proven in general learning settings -- also hold for black-box neural network models. Consequently, a neural network's training performance can serve as a direct probe to detect and map these complex transitions.

Understanding the relationship between learning and simulation carries both theoretical and immediate practical implications. From a practical standpoint, the learning paradigm offers a viable path forward when physical access to a quantum system is restricted \cite{kliesch2021theory}. If an observer only has access to measurement data, direct classical simulation of the underlying Hamiltonian or circuit is impossible. This scenario is particularly salient in the era of cloud-hosted quantum computing, where users must certify the computational capabilities of remote hardware using only output samples \cite{hangleiter2023computational, nguyen2024quantum}. Theoretically, establishing a systematic link between these two paradigms can reveal fundamental connections between how classical devices process quantum matter vs. quantum data \cite{aaronson2011computational}. If a reliable mapping exists, researchers could potentially use one paradigm as a proxy for the other's capacity. For instance, quickly assessing the feasibility of classical simulation by first testing a sample-based learning approach, hence saving significant computational resources.

These theoretical separations, however, typically arise from carefully constructed counterexamples in asymptotic limits. Whether such gaps manifest in realistic quantum systems at accessible scales remains an open empirical question, which we address in this work.

The most direct precedent is the work of Niu \emph{et al.} \cite{niu2020learnability}, who trained deep generative models on random quantum circuits with varying depth and noise levels and quantified complexity through the number of model parameters required to reach a target error. Our setup shares their core approach but differs in the complexity quantification.  Parameter counting, though intuitive, is known to underestimate effective model capacity, since networks with similar parameter counts can differ substantially in expressivity and generalization \cite{maddox2020rethinking}; we therefore characterize learning difficulty through geometric properties of the loss landscape rather than through raw parameter budgets \cite{li2018measuring, ghorbani2019investigation, li2018measuring}.

We fix a single classical learning architecture -- a deep energy-based generative model -- and ask how its optimization landscape and capacity requirements respond as we tune the quantum resource responsible for classical simulation hardness.  Two resources are studied independently:

\begin{enumerate}
  \item \textbf{Entanglement}, varied via the bond dimension $\chi$ of random matrix product states (MPS).  Classical simulation cost in the MPS regime scales polynomially on $\chi$ \cite{schollwock2011density}.
  \item \textbf{Non-stabilizerness}, varied via the number $t$ of $T$ gates in Clifford-dominated circuits. Stabilizer-rank simulation cost scales exponentially on $t$  \cite{bravyi2016improved}.
\end{enumerate}

We characterize learning difficulty with two complementary probes borrowed from the deep-learning literature: the maximum eigenvalue $\lmax$ of the loss Hessian at the converged minimum \cite{ghorbani2019investigation, sagun2017empirical} -- a good proxy for generalization --, and Random Subspace Optimization (RSO) \cite{li2018measuring}, which restricts optimization to a $d$-dimensional random affine subspace of parameter space and thus provides a capacity-constrained measure of representational demand.

Our main finding is that the mapping between simulability and learnability is consistent across the two resources studied. For MPS, both probes agree with the simulation-cost intuition: increasing $\chi$ produces sharper minima and larger reconstruction error under fixed capacity. For Clifford$+T$ circuits the two probes also agree: $\lmax$ grows with the $T$-count, and the RSO-constrained reconstruction error grows with $t$ over a low-$T$ regime before saturating. Both quantum resources therefore manifest in the classical training dynamics and convergence behavior of the neural network under the metrics employed here.

The remainder of the paper is organized as follows. Section \ref{sec:framework} formalizes the simulation/learning comparison and the two resource knobs. Section \ref{sec:methods} describes the model, the training protocol, and the two complexity probes. Section \ref{sec:results} presents the MPS and Clifford$+T$ results. Section \ref{sec:discussion} discusses possible origins of the observed decoupling, and Section \ref{sec:conclusion} concludes.

% ============================================================
\section{Framework}\label{sec:framework}
% ============================================================

\subsection{Simulation versus learning}
A \emph{simulation} algorithm receives a classical description of a quantum system -- a Hamiltonian, a circuit, or a tensor network -- and outputs either samples from the Born distribution $P(x) = |\langle x|\psi\rangle|^2$ (\emph{weak} simulation) or the values of $P(x)$ themselves (\emph{strong} simulation) \cite{nest2008classical}. A \emph{learning} algorithm, in contrast, receives only a finite dataset $\mathcal{D}=\{x^{(1)},\dots,x^{(N_s)}\}$ of measurement outcomes drawn from $P(x)$ and must produce a hypothesis $q_\theta(x)$ that, with high probability, approximates $P$ in some chosen metric \cite{aaronson2007learnability}.  Because both methods produce probabilistic predictors over the same outcome space, they can be compared on identical tasks.

This work uses the learning paradigm as the workbench: a single neural network is trained to reconstruct $P(x)$ from measurement samples, and its training behavior is studied as the underlying physical system becomes harder to be classically simulated.

\subsection{Resource modulators}\label{sec:knobs}

To probe distinct mechanisms of simulation hardness we use two families of states, each with a tunable parameter that increases classical simulation cost while leaving the system size fixed at $N=10$ qubits.

\paragraph*{Matrix product states.}  An $N$-site pure state can be expressed in the MPS form
\begin{equation}
\ket{\psi}= \sum_{i_1,\ldots,i_N=1}^{d} \sum_{\alpha_1,\ldots,\alpha_{N-1}}
A^{[1]\,i_1}_{\alpha_1} A^{[2]\,i_2}_{\alpha_1\alpha_2} \cdots A^{[N]\,i_N}_{\alpha_{N-1}}
\ket{i_1 i_2 \cdots i_N},
\end{equation}

with bond dimension $\chi = \max_i \dim(\alpha_i)$ \cite{schollwock2011density}. The Schmidt rank across any bipartition is bounded by $\chi$, so $\chi$ directly controls entanglement and the computational cost of classical contraction.

\paragraph*{Clifford$+T$ circuits.}  Clifford circuits are classically simulatable in polynomial time via the Gottesman-Knill theorem \cite{gottesman1998heisenberg, aaronson2004improved} despite generating volume-law entanglement.  Injecting non-Clifford resources, typically $T$ gates, is necessary for universal quantum computation \cite{bravyi2005universal}. In the stabilizer-rank formalism, classical simulation cost scales as $\mathcal{O}(2^{c t})$ \cite{bravyi2016improved}, making the $T$-count $t$ an exponential complexity knob.

The two above knobs probe two physical resources. Entanglement is bounded in MPS; non-stabilizerness is bounded by $T$-count in Clifford-dominated circuits. These knobs, however, are not completely independent from each other. Although non-stabilizerness, also known as ``magic'' \cite{bravyi2005universal}, is also present in random MPS \cite{chen2024magic} and hard to properly isolate, we can think of MPS as representatives of bounded-depth quantum circuits \cite{vidal2003efficient} with increasing entanglement, tuned by $\chi$. The same happens for the Clifford+T dataset: even a single $T$ gate drives a transition in the entanglement spectrum statistics \cite{zhou2020single}, and more generally the interplay between entanglement and non-stabilizerness is highly non-trivial -- they are strongly dependent yet statistically uncorrelated \cite{iannotti2025entanglement}, making it difficult to vary one resource while keeping the other fixed.

% ============================================================
\section{Methods}\label{sec:methods}
% ============================================================

\subsection{Dataset generation}

For each value of the resource parameter ($\chi$ for MPS, $t$ for Clifford$+T$) we generate $20$ independent random instances with distinct random seeds and report metrics averaged over these realizations.

\paragraph*{Random MPS.} States are generated with the \texttt{quimb} library \cite{gray2018quimb}.  Local tensors are drawn with i.i.d. standard-normal entries and the chain is brought to mixed canonical form.  This procedure does not sample from the Haar measure on MPS \cite{garnerone2010typicality, chen2024magic}, but produces a narrow entanglement distribution that grows with $\chi$.  The full state vector is obtained by exact contraction and exact-sampled in the computational basis.

\paragraph*{Random Clifford$+T$ circuits.}
Circuits are built with fixed depth $d_{\mathrm{circ}}=500$ in a two-step procedure.  First, $t$ spacetime locations are reserved for $T$ gates by uniform sampling over the $N\times d_{\mathrm{circ}}$ grid.  Second, the remaining slots are filled, layer by layer, by uniformly drawing gates from $\{H,S,\mathrm{CNOT},X,Y,Z,\mathrm{CZ}, \mathrm{SWAP}\}$ and placing them on free qubits within the layer, and then simulated using the \texttt{Qiskit} library \cite{javadi2024quantum}. The large depth saturates entanglement across the explored range of $t$, so that the $T$-count, rather than entanglement, is the dominant complexity knob.

The entanglement profile in both systems are discussed in the Appendix \ref{app:entanglement} in view of each knob's goal: bond dimension should increase entanglement, while adding $t$ gate should not.

\subsection{Energy-based model}

The learner is a fully connected energy network $E_\theta:\{0,1\}^N \to \mathbb{R}$ with five hidden layers of $128$ ReLU units and is trained using Maximum Likelihood Estimation (MLE).  The induced probability distribution is
\begin{equation}\label{eq:loss}
  q_\theta(x) = \frac{e^{-E_\theta(x)}}{Z_\theta}, \qquad
  Z_\theta = \sum_{x\in\{0,1\}^N} e^{-E_\theta(x)}.
\end{equation}

In the studied scale, the partition function $Z_\theta$ is computed exactly at every step, eliminating MCMC noise that otherwise perturbs energy-based training \cite{hinton2002training}. Training uses $N_s=10^5$ samples, Adam with initial learning rate $10^{-4}$, and a reduce-on-plateau scheduler (factor $0.5$, patience $5$). Standard training is run for $200$ epochs. The neural network architecture and training hyperparameters were chosen in a way such that the model converges to the goal distribution in both datasets during the training procedure. This allow us to factor out effects from different networks and compare different datasets faithfully. 

\subsection{Complexity probes}\label{sec:probes}

\paragraph*{Local curvature via the Hessian.}
At the converged parameters $\hat\theta$ we evaluate the Hessian $H(\hat\theta) = \nabla^2_\theta \mathcal{L}(\hat\theta)$ of the full loss
\begin{equation}
  \mathcal{L}(\theta) = \frac{1}{2^N}\sum_{x\in\{0,1\}^N}
                       \ell(x;\theta),
\end{equation}
where $\ell(x;\theta) = -\log q_\theta(x)$ is the loss function employed during training.  Since the partition function in equation (\ref{eq:loss}) is tractable at the studied scale, $H$ is computed over all configurations, yielding the exact Hessian of the full objective with no batch noise. We extract the leading eigenvalue $\lmax$ by power iteration. Sharp minima (large $\lmax$) are empirically associated with harder optimization and poorer generalization \cite{keskar2016large, foret2020sharpness, ghorbani2019investigation}. In a generative setting, generalization does not reduce to a single scalar metric, but it is naturally expressed as the capacity of the model to learn an efficient, compact representation of the data distribution rather than memorizing individual samples \cite{theis2015note}.

The complete Hessian spectrum has shown further properties concerning task complexity. It is composed by two parts: a dense bulk of eigenvalues concentrated with a small number of significant discrete outliers that dominate the training dynamics \cite{sagun2017empirical, sagun2016eigenvalues}. During training, the gradient updates are supported on the subspace spanned by these outliers eigenvectors \cite{gur2018gradient, fort2019emergent}, directly connecting the curvature with the problem of good parameter search. Controlled experiments confirm that increasing class overlap -- a proxy for dataset complexity -- systematically amplifies $\lmax$ \cite{sagun2016eigenvalues}, while optimization algorithms that bias the trajectory toward flatter regions consistently improve generalization \cite{foret2020sharpness, chaudhari2019entropy}.

Sharpness is also connected to sparsity and compression. Because the Hessian governs the second-order expansion of the loss, parameters lying in flat directions can be pruned with negligible degradation \cite{lecun1989optimal}, even in large-scale modern models \cite{frantar2023sparsegpt}. Simpler problems therefore admit solutions in flatter regions of parameter space, requiring fewer effective degrees of freedom.

Finally, the Hessian spectrum offers a geometric interpretation of the constraints imposed by the data. The eigenvectors associated with the largest eigenvalues identify the directions in parameter space most tightly constrained by the training distribution as perturbations along these directions produce the sharpest change in the loss \cite{koh2017understanding}. Strikingly, this geometric transition is observable dynamically as models that undergo grokking, a delayed generalization phase following apparent convergence, exhibit a marked drop in sharpness precisely at the moment generalization occurs \cite{power2022grokking, han2026flatness}. In this sense, $\lmax$ and the outlier spectrum reflect how many and how strongly the data bind the model, providing a direct signature of dataset complexity.

\paragraph*{Capacity-constrained training via RSO.}
Following Li et al. \cite{li2018measuring}, we restrict optimization to a random $d$-dimensional affine subspace of parameter space:
\begin{equation}
  \theta^{(D)} = \theta^{(D)}_0 + P\,\theta^{(d)},
\end{equation}
with $\theta_0^{(D)}$ the random initial point, $P\in\mathbb{R}^{D\times d}$ a fixed random isometry, and $\theta^{(d)}\in\mathbb{R}^d$ the only trainable variable. The original intrinsic dimension measure seeks the smallest $d$ that recovers $90\%$ of full-model performance, $\dint$; in some tasks, $d$ is as small as $0.4\%$ of the full parameter space. This extreme level of compression is related to the manifold hypothesis \cite{bengio2013representation}, which states that data lives in a low-dimensional manifold that is learned by well trained networks, and inspired prominent works on neural network compression and efficient fine-tuning \cite{hu2022lora, dettmers2022gpt3}.

Similar to this procedure is the Lottery Ticket Hypothesis (LTH) \cite{frankle2018lottery}, which identifies a sparse subnetwork capable of matching full-model performance when trained in isolation. While RSO compresses via low-rank optimization, LTH achieves this through pruning --- and this structural approach recently yielded promising results in Neural Quantum States. There, magnitude-based pruning revealed that the optimal subnetwork sparsity and error-scaling behavior adapt dynamically to the physical phase of the target state, allowing structurally simpler phases to be represented with significantly fewer connections \cite{barton2026connectivity}.

Here we use RSO as a capacity constraint rather than as a direct $\dint$ estimator. Concretely, we fix $d$ and ask how the converged reconstruction error responds to the resource parameter, essentially fixing the classical budget available for learning the correlations in data, enabling fair cross-distribution comparison. 

We also explored varying the network capacity through standard architectural changes -- adding layers or increasing hidden-layer width -- as an alternative way to probe learnability. However, such discrete modifications introduced substantial noise into the results: performance changes were abrupt and incommensurate across configurations, failing to resolve the smooth complexity gradients present at the scales explored here. This is consistent with the known non-linear scaling of expressive power with depth and width \cite{raghu2017expressive}, which makes architectural parameters poor proxies for a continuous capacity budget. The RSO subspace dimension $d$, by contrast, provides a single continuous axis that modulates representational capacity without altering the network topology \cite{aghajanyan2021intrinsic}, enabling a fair and smooth comparison across distributions of varying quantum resource content.

Reconstruction quality is then measured by the total-variation distance $\dtv = \tfrac12 \sum_x |P(x)-q_\theta(x)|$, which is both a lower bound on the trace distance \cite{nielsen2010quantum} and an upper bound on the KL divergence by Pinsker's inequality \cite{cover1999elements}, meaning it is also being minimized by the MLE-based training procedure.

% ============================================================
\section{Results}\label{sec:results}
% ============================================================

\subsection{Entanglement: MPS}\label{sec:mps_results}

\begin{figure}[ht]
  \centering
  \includegraphics[width=\columnwidth]{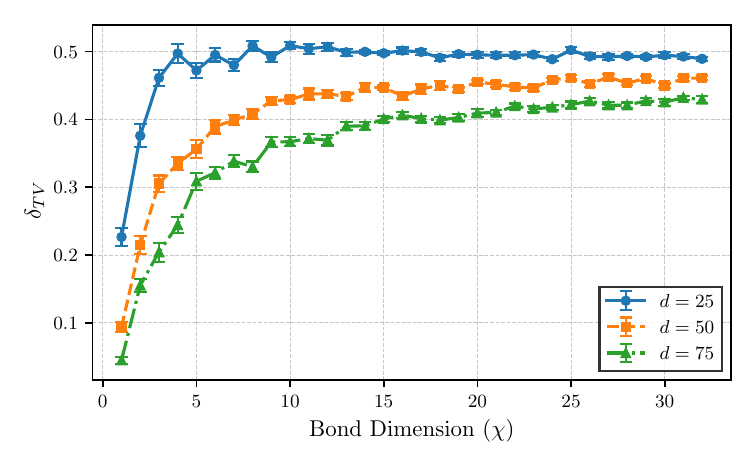}
  \caption{\textbf{RSO error versus bond dimension for random MPS.} Average total-variation distance $\dtv$ between the learned and target distributions, as a function of bond dimension $\chi$, for three optimization-subspace dimensions $d\in\{25,50,75\}$.  At fixed $d$, error grows with $\chi$; at fixed $\chi$, error decreases with $d$.  Each point averages $20$ random MPS instances; error bars denote one standard deviation.}
  \label{fig:mps_rso}
\end{figure}

\begin{figure}[ht]
  \centering
  \includegraphics[width=\columnwidth]{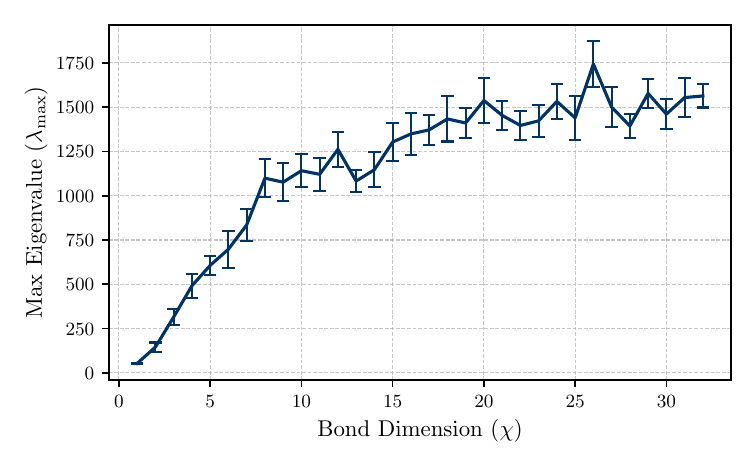}
  \caption{\textbf{Maximum Hessian eigenvalue versus bond dimension.} $\lmax$ at convergence grows almost monotonically with $\chi$, indicating that higher entanglement drives the optimization landscape towards sharper minima.}
  \label{fig:mps_hessian}
\end{figure}

For random MPS, both probes respond consistently to the entanglement knob. Figure \ref{fig:mps_rso} shows that, at any fixed optimization subspace dimension $d$, the reconstruction error $\dtv$ increases as $\chi$ grows: more entangled targets are harder to represent with the same available capacity. Conversely, at fixed $\chi$ the error decreases as $d$ grows, confirming that the constraint behaves as a fair capacity bottleneck. In the most restrictive regime ($d=25$) the error saturates near $0.5$ for $\chi\gtrsim 8$, suggesting that the available degrees of freedom are insufficient to resolve differences between targets of varying entanglement. Indeed, as we decrease the subspace dimension $d$, we impose more constraints between the model parameters.  In this scenario, after finding a good local minimum, the optimizer cannot distribute the probability mass in a better way, because it would obligatory change the probability of other bitstrings in a unintended way. Because of that, only sufficient higher $d$ restores resolution and exposes the monotonic dependence on $\chi$. This is an important prerequisite of this method and further experiments in this matter are discussed at Appendix \ref{app:rso_convergence}.

Figure \ref{fig:mps_hessian} shows that $\lmax$ at convergence also grows monotonically with $\chi$. Note that all data points were obtained using the same architecture and training configuration, differing only on the dataset. Hence this is not a generic feature of neural-network optimization. Standard initialization schemes place all eigenvalues close to zero, and the outliers reported here emerge only as the network adapts to the data \cite{sagun2017empirical}. 

Taken together, the two probes indicate that for MPS the increase in classical simulation cost is reflected both in local landscape sharpness and in global representational demand.

\subsection{Non-stabilizerness: Clifford+T circuits}\label{sec:clifford_results}

\begin{figure}[ht]
  \centering
  \includegraphics[width=\columnwidth]{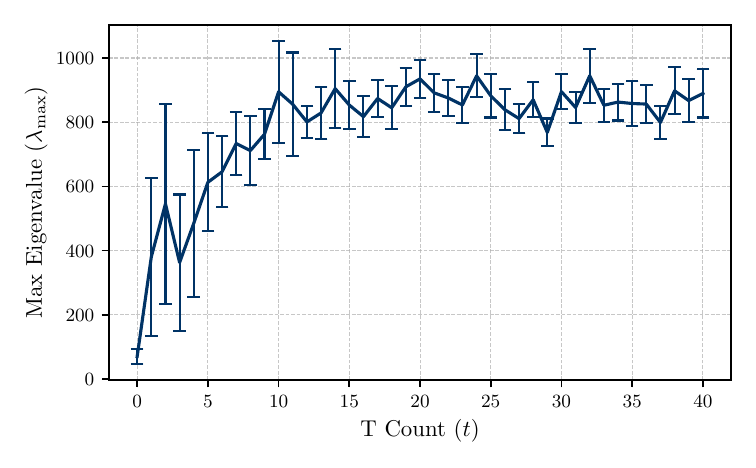}
  \caption{\textbf{Maximum Hessian eigenvalue versus $T$-count.} $\lmax$ at convergence increases with $t$ but exhibits substantial instance-to-instance variability for low $T$-counts, reflecting the heterogeneity of Clifford$+T$ Born distributions in the computational basis. Moreover, there is a saturation at $t=10$, indicating that a maximum of quantum resources was attained at that point.}
  \label{fig:clifford_hessian}
\end{figure}

\begin{figure}[ht]
  \centering
  \includegraphics[width=\columnwidth]{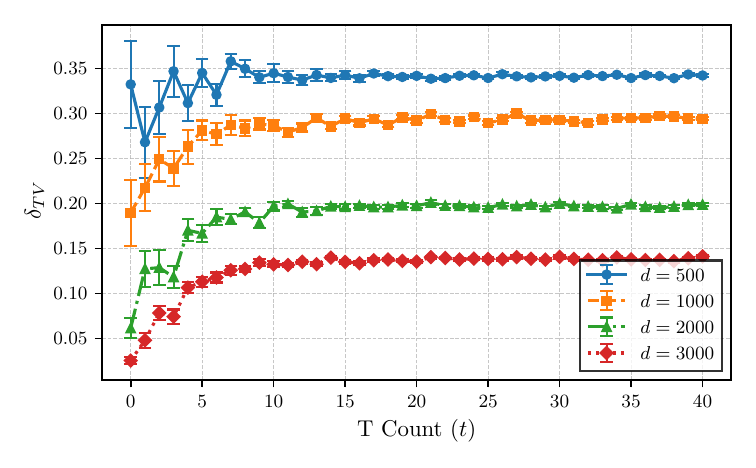}
  \caption{\textbf{RSO error versus $T$-count.}
  Average $\dtv$ as a function of $t$ for a fixed subspace dimension. There is an error increasing pattern until saturation at $t=10$.}
  \label{fig:clifford_rso}
\end{figure}

Clifford$+T$ circuits showed similar results. Figure \ref{fig:clifford_hessian} shows that $\lmax$ trends upwards with the $T$-count: injecting non-stabilizer gates produces sharper minima in the trained network. The trend is noisier than for the MPS case, an effect we attribute to the strong instance-to-instance variability of Clifford$+T$ Born distributions in the computational basis, which range from highly concentrated supports for low $t$ to broader distributions for larger $t$. 
Despite the noise, the average local curvature increases with $t$, consistent with the qualitative expectation from the resource theory of non-stabilizerness \cite{veitch2014resource}. 

Similarly, Fig. \ref{fig:clifford_rso} also shows a monotonic dependence of the RSO reconstruction error on $t$ for $d\in\{500, 1000, 2000, 3000\}$ up to a saturation point at $t\approx 10$.  
We note that this monotonic signal only emerges at sufficiently large subspace dimensions, and at smaller values, no clear trend is visible (see Appendix \ref{app:rso_convergence}). We attribute this to the near-maximal entanglement already present at $t=0$. When the base entanglement saturates the model's available capacity, the subtler effect of magic is obscured. Only once $d$ is large enough to accommodate the entanglement structure, the additional complexity from non-stabilizerness is apparent.

% ============================================================
\section{Discussion}\label{sec:discussion}
% ============================================================

The results indicate that the relationship between simulability and learnability is consistent across the two resources studied, though with quantitative differences. In the MPS setting, increasing entanglement affected both restricted capacity error and loss curvature monotonically across the entire range of $\chi$ explored.  Simulation complexity and learning complexity appear aligned in this regime: the same structural constraint that challenges tensor network contraction also manifests as geometric stiffness in the neural network's optimization landscape.

Both metrics respond to non-stabilizerness in the same  qualitative way as they respond to entanglement, with the difference  that the Clifford$+T$ response saturates within the explored range  while the MPS response does not. This saturation is consistent with the behavior of magic measures under random Clifford$+T$ circuits. Beyond a critical T-count, the non-stabilizerness of the state itself  saturates, and additional T gates no longer increase the statistical complexity of the resulting Born distribution \cite{zhou2020single, turkeshi2025magic, veitch2014resource}.

The requirement for large $d$ in the Clifford$+T$ case, but not in the MPS case, also has a physical interpretation. In the MPS setting, entanglement is the only varying resource, so any available capacity is directly sensitive to it. In the Clifford$+T$ setting, entanglement is saturated at its maximum value even at $t=0$ (see Appendix~\ref{app:entanglement}), meaning the model must first represent a highly entangled distribution before it can resolve the additional structure introduced by magic. At small $d$, the entanglement complexity alone exhausts the available degrees of freedom, leaving no residual capacity to detect the non-stabilizerness signal. This suggests a hierarchy where entanglement sets a baseline representational demand that must be met before magic becomes visible to the learner.

% ============================================================
\section{Conclusion and open questions}\label{sec:conclusion}
% ============================================================

We investigated whether classical simulation hardness implies classical learning hardness for quantum probability distributions. Using a deep energy-based generative model, we systematically varied quantum resources and measured their impact on neural network optimization geometry and restricted-capacity performance.
 
For Matrix Product States and Clifford+$T$ circuits, we observed a positive correlation between the quantum resources -- entanglement and magic, respectively -- and learning complexity. As $\chi$ and $t$ increased, both the reconstruction error under fixed RSO subspace dimension and the maximum Hessian eigenvalue grew monotonically across the range explored.

However, for the Clifford+$T$ circuit, the saturation of both probes at $t \approx 10$ distinguishes this case from the MPS sweep, and is consistent with the saturation of magic as a property of the target states. Verifying this interpretation would require running the same protocol at larger system sizes, where the saturation threshold is expected to shift \cite{leone2021quantum}.

Our geometric analysis relied solely on the largest Hessian eigenvalue due to computational constraints. While $\lambda_{\max}$ captures the sharpest curvature direction, the full Hessian eigenspectrum contains richer information about task structure \cite{ghorbani2019investigation, maddox2020rethinking}. Stochastic Lanczos methods \cite{ghorbani2019investigation, yao2021adahessian} make it feasible to approximate the full spectrum, and applying such techniques could reveal finer correlations not visible through $\lambda_{\max}$ alone.

The RSO results suggest a natural extension. Here we used RSO as a capacity constraint at fixed $d$. A direct estimate of the intrinsic dimension $\dint$ in the sense of Ref. \cite{li2018measuring} -- the smallest $d$ that recovers a target fraction of full-model performance -- would provide a single number per dataset that aims to be consistent across architectures, and would allow a more direct comparison between the two resources.

A further limitation concerns the measurement basis. Our model is trained exclusively on Born distributions in the computational basis, while magic is fundamentally a coherent resource that manifests through phase correlations and interference structure. Single-basis sampling may therefore only partially capture the complexity introduced by non-stabilizerness, and the signals observed here may underestimate the true effect. Measurements in multiple bases, or access to off-diagonal elements of the density matrix, could reveal stronger and cleaner correlations between T-count and learning difficulty.
 
The empirical protocol introduced here can also be generalized to additional quantum resources. Beyond entanglement and non-stabilizerness, one could investigate Bell nonlocality \cite{brunner2014bell}, steering \cite{uola2020quantum}, or other quantum correlation measures \cite{bera2018quantum, horodecki2009quantum} as tunable parameters within comparable experimental designs. Likewise, alternative learning metrics such as the intrinsic dimension of activation manifolds \cite{facco2017estimating}, information-theoretic complexity measures \cite{weimar2025fisher}, or diffusion-model-based estimators \cite{kamkari2024geometric, tempczyk2022lidl} may reveal different aspects of how quantum structure manifests in classical optimization.
 
Beyond the specific question of simulability versus learnability, the framework developed here may serve as a practical tool for model comparison \cite{weimar2025fisher}. Learning complexity metrics such as the ones used here provide signals about how a given architecture perceives the structure of a dataset, and could guide architectural choice in variational settings such as Neural Quantum States \cite{passetti2023can, denis2025comment}, reinforcement learning for quantum control \cite{bukov2018reinforcement}, or quantum error correction decoding \cite{valenti2019hamiltonian}.

% \begin{acknowledgments}

% The authors thank Alexandre Baron Tacla, Daniel Brod, Dario Rosa, and for fruitful discussions and valuable suggestions.

% This work is supported in part by the National Council for Scientific and Technological Development project 
% CNPq Brazil (project: Universal Grant No. 408990/2025-2). FdM also acknowledges financial support from CNPq projects 305071/2022-0, (JP) 408153/2022-9 and , the Carlos Chagas Foundation for Research Support of the State of Rio de Janeiro (FAPERJ, Grant APQ1 E-26/210.576/2024), and it is part of the National Institute of Science and Technology for Applied Quantum Computing through CNPq process No. 408884/2024-0.
% \end{acknowledgments}

\begin{acknowledgments}
The authors thank Alexandre Baron Tacla, Daniel Brod and Dario Rosa for fruitful discussions and valuable suggestions.
This work is supported in part by the National Council for Scientific and Technological Development, 
CNPq Brazil (Universal Grant No. 408990/2025-2), the Carlos Chagas Foundation for Research Support of the State of Rio de Janeiro (FAPERJ, Grant APQ1 E-26/210.576/2024), and it is part of the National Institute of Science and Technology for Applied Quantum Computing through CNPq process No. 408884/2024-0. FdM also acknowledges financial support by CNPq (grant: 305071/2022-0), and fellowship by EMBRAPII Senai-CIMATEC - QuIIN (Quantum Industrial Innovation). JPdR acknowledges partial funding from CNPq project 408153/2022-9.
\end{acknowledgments}

\appendix

\section{Entanglement concentration in both datasets}\label{app:entanglement}

A central assumption of our experimental setup is that both knobs operate as intended. For MPS, bond dimension $\chi$ should directly control entanglement, while for Clifford+$T$ circuits, the large circuit depth $d_{\mathrm{circ}}=500$ should saturate entanglement across all explored values of T-count $t$.

\begin{figure}[!ht]
    \centering
    \includegraphics[width=\linewidth]{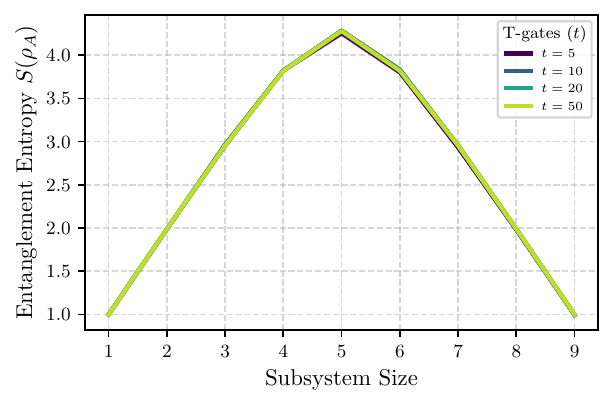}
    \caption{Entanglement entropy $S(\rho_A)$ as a function of subsystem  size for Clifford$+T$ circuits at depths $d_{\mathrm{circ}}=500$, for T-counts $t \in \{5, 10, 20, 50\}$. All curves collapse onto a single volume-law profile, confirming that entanglement is saturated across the full range of T-counts explored and is therefore not the driver of the complexity signal observed in the main text.}
    \label{fig:entanglement_clifford}
\end{figure}

\begin{figure}[!ht]
    \centering
    \includegraphics[width=\linewidth]{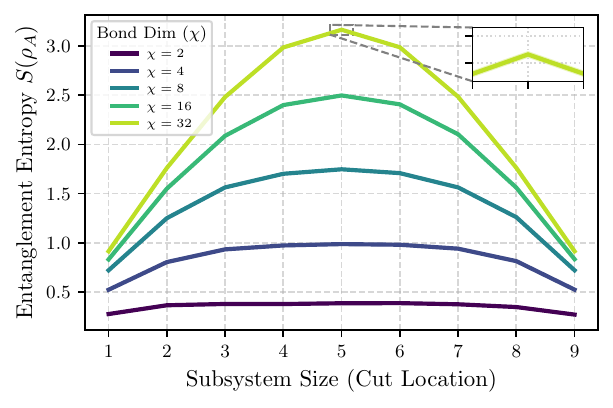}
    \caption{Entanglement entropy $S(\rho_A)$ as a function of bipartition  cut location for random MPS at bond dimensions $\chi \in \{2, 4, 8, 16, 32\}$. Entropy grows systematically with $\chi$, confirming that  bond dimension acts as an effective entanglement knob across the explored range. The inset displays a close-up of the error bands, which are smaller than the line width.}
    \label{fig:entanglement_mps}
\end{figure}

Figures \ref{fig:entanglement_clifford} and \ref{fig:entanglement_mps} confirm both assumptions. For Clifford+$T$ circuits, the entanglement entropy as a function of subsystem size is nearly identical throughout different values of $t$, exhibiting the expected volume law profile, and confirming that entanglement, under this measure, is held constant. For random MPS, as expected, the entropy for a fixed subsystem increases with the bond dimension $\chi$, confirming our experimental setup.

\section{Convergence behavior under RSO capacity constraints}\label{app:rso_convergence}

A potential concern with the RSO results in the main text is that the models trained in low-dimensional subspaces may simply be under-trained. If the optimizer has not converged, the measured $\dtv$ reflects insufficient training time rather than a true capacity limitation.

\begin{figure}[t]
    \centering
    \includegraphics[width=\linewidth]{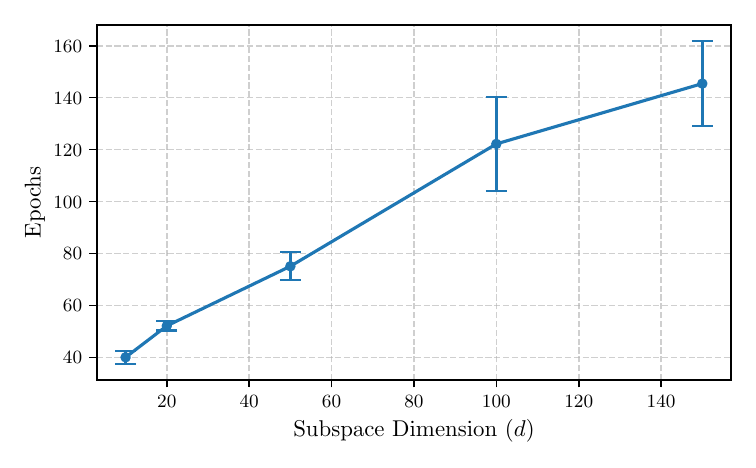}
    \caption{Number of training epochs required for convergence as a function of RSO subspace dimension $d$, for the MPS dataset at $\chi=16$. Convergence is determined by early stopping with patience $5$. The monotonic trend confirms that low $d$ models are not under trained. They converge faster but to a larger error, reflecting a capacity bottleneck rather than insufficient optimization time.}
    \label{fig:epoch_vs_dim}
\end{figure}

Figure \ref{fig:epoch_vs_dim} addresses this directly. We trained models without a fixed epoch limit, using early stopping with a patience of $5$ epochs without improvement, and recorded the total epochs required for convergence as a function of subspace dimension $d$. As shown, larger subspaces required more training epochs. Therefore, models at small $d$ do not attain higher $\dtv$ due to lack of training, but reach the best solution their restricted capacity allows. As discussed in the main text, this is due to the lack of freedom to independently reorganize the probability amplitudes to fit the objective distribution, imposed by the constraints on gradient updates.

\begin{figure}[t]
    \centering
    \includegraphics[width=\linewidth]{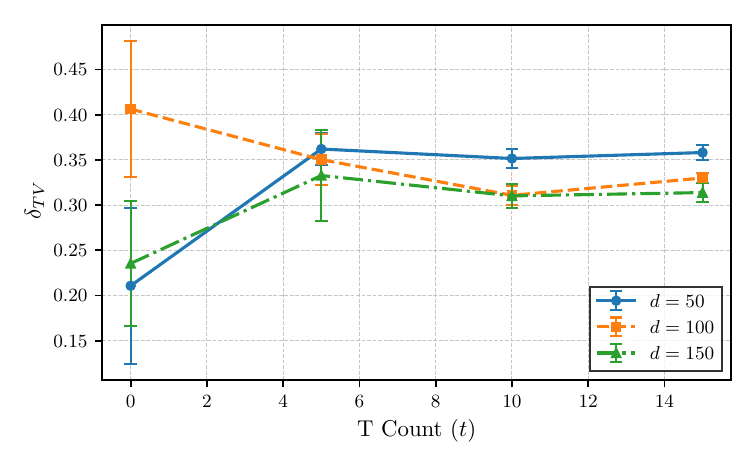}
    \caption{RSO reconstruction error $\dtv$ as a function of T-count for subspace dimensions $d \in \{50, 100, 150\}$. At these capacity levels, no monotonic trend is visible and  curves for different $d$ largely overlap, indicating that  the model is bottlenecked by the base entanglement of the  circuits. The magic signal only becomes resolvable at  significantly larger $d$ (see Figure \ref{fig:clifford_rso} in the main text),  demonstrating that sufficient representational capacity  is a prerequisite for detecting complexity increase through the RSO probe.}
    \label{fig:clifford_neg_result}
\end{figure}

Figure \ref{fig:clifford_neg_result} further illustrates this point for the Clifford+$T$ dataset. At subspace dimensions comparable to the MPS experiments ($d \in \{50, 100, 150\}$), the reconstruction error shows no clear monotonic trend with T-count, and the curves for different $d$ are largely overlapping. This result is significantly different from Figure \ref{fig:clifford_rso} at the main text, where a clear monotonic signal is present. The comparison demonstrates that a minimum capacity is a prerequisite for resolving the complexity gradient induced by quantum resources. At small $d$, the model is bottlenecked by the base entanglement of the circuit regardless of T-count, and the magic influence is obscured.

\newpage

\bibliographystyle{apsrev4-2}
\bibliography{ref.bib}

\end{document}